\documentclass[preprint]{aastex63}

\newcommand{\lya}{Ly$\alpha$}

\shorttitle{the flaring \lya\ emission observed by SO/EUI}
\shortauthors{Li et al.}

\begin{document}

\title{The Lyman-$\alpha$ Emission in a C1.4 Solar Flare Observed by the Extreme Ultraviolet Imager aboard Solar Orbiter}

\author{Ying Li} 
\author{Qiao Li}
\author{De-Chao Song}
\affiliation{Key Laboratory of Dark Matter and Space Astronomy, Purple Mountain Observatory, Chinese Academy of Sciences, Nanjing 210023, People’s Republic of China; yingli@pmo.ac.cn}
\affiliation{School of Astronomy and Space Science, University of Science and Technology of China, Hefei 230026, People’s Republic of China}

\author{Andrea Francesco Battaglia}
\affiliation{University of Applied Sciences and Arts Northwestern Switzerland, Bahnhofstrasse 6, 5210 Windisch, Switzerland}
\affiliation{ETH Z{\"u}rich, R{\"a}mistrasse 101, 8092 Z{\"u}rich, Switzerland}

\author{Hualin Xiao}
\affiliation{University of Applied Sciences and Arts Northwestern Switzerland, Bahnhofstrasse 6, 5210 Windisch, Switzerland}

\author{S{\"a}m Krucker}
\affiliation{University of Applied Sciences and Arts Northwestern Switzerland, Bahnhofstrasse 6, 5210 Windisch, Switzerland}
\affiliation{Space Sciences Laboratory, University of California, 7 Gauss Way, 94720 Berkeley, USA}

\author{Udo Sch{\"u}hle}
\affiliation{Max Planck Institute for Solar System Research, Justus-von-Liebig-Weg 3, 37077 G{\"o}ttingen, Germany}

\author{Hui Li}
\author{Weiqun Gan}
\affiliation{Key Laboratory of Dark Matter and Space Astronomy, Purple Mountain Observatory, Chinese Academy of Sciences, Nanjing 210023, People’s Republic of China; yingli@pmo.ac.cn}
\affiliation{School of Astronomy and Space Science, University of Science and Technology of China, Hefei 230026, People’s Republic of China}

\author{M. D. Ding}
\affiliation{School of Astronomy and Space Science, Nanjing University, Nanjing 210023, People’s Republic of China}
\affiliation{Key Laboratory for Modern Astronomy and Astrophysics (Nanjing University), Ministry of Education, Nanjing 210023, People’s Republic of China}


\begin{abstract}
The hydrogen Lyman-$\alpha$ (H {\sc i} Ly$\alpha$) emission during solar flares has rarely been studied in spatially resolved images and its physical origin has not been fully understood. In this paper, we present novel Ly$\alpha$ images for a C1.4 solar flare (SOL2021-08-20T22:00) from the Extreme Ultraviolet Imager aboard Solar Orbiter, together with multi-waveband and multi-perspective observations from the Solar Terrestrial Relations Observatory Ahead and the Solar Dynamics Observatory spacecraft. It is found that the Ly$\alpha$ emission has a good temporal correlation with the thermal emissions at 1--8 \AA\ and 5--7 keV, indicating that the flaring Ly$\alpha$ is mainly produced by a thermal process in this small event. However, nonthermal electrons play a minor role in generating Ly$\alpha$ at flare ribbons during the rise phase of the flare, as revealed by the hard X-ray imaging and spectral fitting. Besides originating from flare ribbons, the Ly$\alpha$ emission can come from flare loops, likely caused by plasma heating and also cooling that happen in different flare phases. It is also found that the Ly$\alpha$ emission shows fairly similar features with the He {\sc ii} 304 \AA\ emission in light curve and spatio-temporal variation along with small differences. These observational results improve our understanding of the Ly$\alpha$ emission in solar flares and also provide some insights for investigating the Ly$\alpha$ emission in stellar flares.
\end{abstract}


\section{Introduction}
\label{sec:intro}

The hydrogen Lyman-$\alpha$ (H {\sc i} \lya) line at 1216 \AA\ is the strongest line among the solar ultraviolet (UV) spectra \citep{curd01}. This line is optically thick \citep{vial82,wood95} with a central reversal in the line core and formed mostly in the chromosphere and transition region \citep[e.g.,][]{vern81,vour10}. It has been found that during a solar flare, the \lya\ line emission can be notably enhanced \citep[e.g.,][]{wood04,mill12,kret15,lule21}. However, its physical origin and property have not been fully understood, due to a lack of spatially resolved images in flares. In addition, the \lya\ emission from other stars is usually absorbed by the interstellar medium and cannot be directly observed from Earth \citep[e.g.,][]{lins13}. Therefore, studying the \lya\ emission in solar flares can also help to understand the stellar flares.

In the past few decades, the flaring \lya\ emission was mainly studied in photometric and spectroscopic observations along with radiative hydrodynamic simulations. \cite{canf80} and \cite{lema84} presented the \lya\ line profiles and their temporal evolution during solar flares observed by Skylab and OSO-8. \cite{wood04} reported that the \lya\ line wing is enhanced by about two times while the line core by 20\% in an X17 flare. Using the \lya\ fluxes of more than 400 X- and M-class flares measured by the Geostationary Operational Environmental Satellite (GOES), \cite{mill20} found that the \lya\ emission has an enhancement of 10\% or less in most of the flares, with an average of 1--4\%, while the enhancement is only about 0.1--0.3\% on average for thousands of C- and B-class flares \citep{mill21}. The physical origin of flaring \lya\ has also been studied by comparing \lya\ with the hard X-ray (HXR) and soft X-ray (SXR) emissions that are associated with nonthermal electrons and thermal plasmas, respectively. \cite{nusi06} showed that the \lya\ emission reaches its maximum before the SXR peak but around the same time with the HXR emission above 50 keV in an X5.3 flare, indicative of a nonthermal origin of \lya. This Neupert effect \citep{neup68} was also detected in some other flares for \lya\  \citep[e.g.,][]{rubi09,mill16,mill17,cham18,domi18,waut22}. Interestingly, \cite{jing20} studied 658 M- and X-class flares and found that in $\sim$20\% of the events the \lya\ peak appears later than the SXR peak. This delayed \lya\ emission can mostly be contributed by thermal plasmas that cool down, suggesting a thermal origin. In fact, both the nonthermal and thermal origins of \lya\ have been investigated and confirmed in flare simulations with different heating models considered \citep{heno95,fang95,fang00,brow18,hong19,yang21}. It should be mentioned that the flaring \lya\ emission was also explored in imaging observations but quite rarely. \cite{rubi09} presented some \lya\ images for an M1.4 flare observed by the Transition Region and Coronal Explorer (TRACE; \citealt{hand99}). They found that most of \lya\ comes from flare footpoints and is co-spatial with HXR sources, also indicating a nonthermal origin. To our knowledge, this is the only imaging observation in \lya\ for flares that has been reported so far.

Launched in 2020, Solar Orbiter (SO; \citealt{mull20}) carries an Extreme Ultraviolet Imager (EUI; \citealt{roch20}) that can provide imaging observations at EUV 174 \AA\ and 304 \AA\ as well as \lya. It also carries a Spectrometer Telescope for Imaging X-rays (STIX; \citealt{kruk20}) observing solar HXR emissions. It is fortunate that EUI has captured solar flares above C-class in \lya\ with simultaneous HXR observations from STIX. The current paper presents probably the first observation of flaring \lya\ from EUI, together with multi-waveband and multi-perspective observations from the EUV Imager (EUVI; \citealt{howa08}) aboard the Solar Terrestrial Relations Observatory Ahead (STEREO-A; \citealt{kais08}) and the Atmospheric Imaging Assembly (AIA; \citealt{leme12}) aboard the Solar Dynamics Observatory (SDO; \citealt{pesn12}). These comprehensive observations can help understand the physical origin and property of the flaring \lya\ emission.

\section{Instruments and Data}
\label{sec:data}

The data we used in this work come from multiple instruments including SO/EUI, SO/STIX, STEREO-A/EUVI, SDO/AIA, SDO/HMI (Helioseismic and Magnetic Imager; \citealt{sche12}), and GOES/XRS (X-Ray Sensor; \citealt{hans96}), which can provide multi-waveband and multi-perspective solar images or light curves. The observing wavebands plus some information on the data are listed in Table \ref{tab:data}.

\begin{table}[htb]
\caption{Instruments and wavebands used in this study along with some information on the data}
\label{tab:data}
\begin{tabular}{lcccccc}
\hline
\hline
Instrument & Waveband & Ion & log($T$) & Cadence & Pixel Size$^{a}$ & Perspective$^{b}$ \\
\hline
SO/EUI & HRI \lya         & H {\sc i} & 4.1--4.7 & 15/5 s & 0.5\arcsec/1.0\arcsec & face view \\
             & FSI 304 \AA   & He {\sc ii} & 4.7 & - & 4.4\arcsec & face view \\
SO/STIX & $>$4 keV    & - & $\ge$6.8 & 1--20 s & - & face view \\
\hline
STEREO-A/EUVI & 304 \AA & He {\sc ii} & 4.7 & 10 min & 1.6\arcsec & face view \\
                             & 195 \AA & Fe {\sc xii}/{\sc xxiv} & 6.2/7.3 & 2.5 min & 1.6\arcsec  & face view \\
\hline
SDO/AIA & 304 \AA & He {\sc ii} & 4.7 & 12 s & 0.6\arcsec & side view \\
               & 193 \AA & Fe {\sc xii}/{\sc xxiv} & 6.2/7.3 & 12 s & 0.6\arcsec & side view \\
               & 211 \AA & Fe {\sc xiv} & 6.3 & 12 s & 0.6\arcsec & side view \\
               & 335 \AA & Fe {\sc xvi} & 6.4 & 12 s & 0.6\arcsec & side view \\
               &  94 \AA & Fe {\sc xviii} & 6.8 & 12 s & 0.6\arcsec & side view \\
               & 131 \AA & Fe {\sc viii}/{\sc xxi} & 5.6/7.0 & 12 s & 0.6\arcsec & side view \\
SDO/HMI & 6173 \AA & Fe {\sc i} & 3.7 & 45 s & 0.5\arcsec & side view \\
\hline
GOES/XRS & 1--8 \AA & - & $\ge$6.6 & 1 s & - & side view \\
\hline
\hline
\end{tabular}
$^{a}$ Note that the unit arc-second of EUI corresponds to $\sim$475 km on the Sun at a distance of 0.65 AU, unlike the other instruments (i.e., EUVI, AIA, and HMI) corresponding to $\sim$725 km at $\sim$1 AU. \\
$^{b}$ SO and STEREO-A had separation angles of ${81.5}^{\circ}$ and ${42.7}^{\circ}$ with the Earth respectively when observing the flare event under study. SDO and GOES observed the event from the Earth perspective.
\end{table}

EUI consists of three telescopes with one Full Sun Imager (FSI) and two High Resolution Imagers (HRIs). FSI provides the full-Sun images at 304 \AA\ and 174 \AA\ while the HRIs observe a partial region of the Sun in \lya\ and 174 \AA\ with a high resolution. Here we mainly use the HRI \lya\ data\footnote{https://doi.org/10.24414/s5da-7e78} of level-2 (L2). The \lya\ images have been rotated north up and aligned among different frames by using a sub-pixel image registration  code ($sir2.pro$; \citealt{feng12}) which is based on cross correlation and has a sub-pixel accuracy. For comparison, the \lya\ images are also co-aligned with EUVI images after derotation and regulation by taking into account the position of spacecraft. This co-alignment has an uncertainty of a few pixels. STIX is a HXR imaging spectrometer observing the Sun in a range of 4--150 keV with energy and angular resolutions of $\sim$1 keV and 7--180\arcsec, respectively. We use the processed light curves of 5--7 keV and 13--18 keV. The HXR images are obtained from L1 pixel data. For both EUI and STIX data, we adopt the light-arrival time at Earth with a correction of 176.4 s for the flare under study. EUVI observes the full Sun in four EUV passbands and we mainly use the images at 304 \AA\ and 195 \AA. AIA obtains the full-Sun images in multiple EUV and UV channels, most of which are used in this work. HMI can provide the full-Sun magnetogram via the Fe {\sc i} line at 6173 \AA. The SXR 1--8 \AA\ data are from GOES-17/XRS for this flare.

\section{Observations and Results}
\label{sec:result}

\subsection{Overview of the Flare Event}

\begin{figure}
\centering
\includegraphics[width=11.5cm]{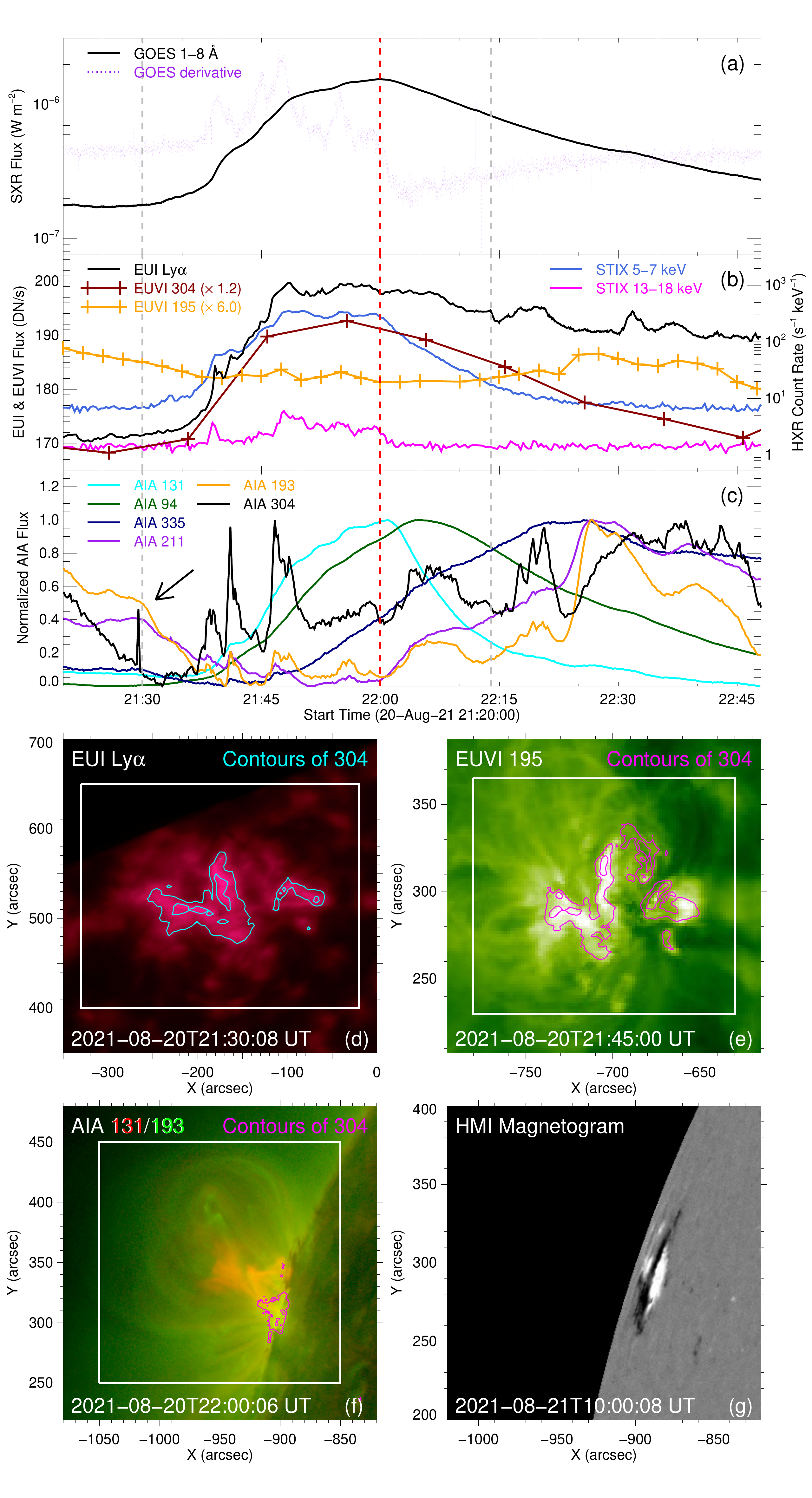}
\caption{(a)--(c) Light curves in multiple wavebands from GOES, STIX, EUI, EUVI, and AIA. The red vertical line denotes the SXR peak time and the two gray lines denote the onset and end times of the flare. The black arrow in panel (c) indicates the coronal rain before the flare onset. (d)--(f) EUI \lya, EUVI 195 \AA, and AIA 131/193 \AA\ images in different flare phases, overlaid with contours of FSI 304 \AA\ (at 21:10:23 UT), EUVI 304 \AA\ (at 21:45:45 UT), and AIA 304 \AA\ (at 22:00:05 UT), respectively. The white box in each panel marks the flaring region used to make the light curves in panels (b) and (c). (g) HMI magnetogram for the active region $\sim$12 hours later.}
\label{fig:lc}
\end{figure}

The event under study is a GOES C1.4 flare (SOL2021-08-20T22:00) that began at 21:30 UT, peaked at 22:00 UT, and ended at 22:14 UT on 2021 August 20 (see the SXR 1--8 \AA\ light curve in Figure \ref{fig:lc}(a)). It was located at the east limb with part of structures occulted by the solar disk from the Earth perspective as seen in the AIA field of view (Figure \ref{fig:lc}(f)). By contrast, EUI and EUVI observed the whole flaring region from another two different perspectives with an angle of 38.8$^{\circ}$ apart (Figures \ref{fig:lc}(d) and (e)). The light curves of EUI \lya, EUVI 304 \AA, and 195 \AA\ from the flaring region (indicated by the white box in the panels) are shown in Figure \ref{fig:lc}(b), together with STIX HXR fluxes. One can see that all these light curves show an obvious enhancement during the flare. Note that the EUVI 195 \AA\ curve looks very similar in the trend with the one of AIA 193 \AA\ as shown in Figure \ref{fig:lc}(c), implying that the flare loops with different length scales are mostly above the limb and their emissions can be mainly captured by AIA in spite of suffering a limb absorption by cool materials such as prominences or spicules in front of the loops. This is confirmed from the AIA image in Figure \ref{fig:lc}(f) that shows hot flare loops (mainly at 131 \AA) plus background coronal loops (mainly at 193 \AA). From the multiple AIA light curves (Figure \ref{fig:lc}(c), integrated over the region marked by the white box in Figure \ref{fig:lc}(f)), we can see that the 131 \AA\ flux reaches its maximum around the same time with the SXR flux and the other EUV fluxes peak later successively exhibiting a typical cooling trend of flare loops. Also note that in this event, multi-thermal coronal rains \citep[e.g.,][]{anto15,scul16,lixi22} can be seen before the flare onset, as shown in the light curves of AIA 211 \AA, 193 \AA, and 304 \AA\ (denoted by the black arrow). Finally, Figure \ref{fig:lc}(g) gives the HMI magnetogram about 12 hours later and we can see complex magnetic polarities that are supposed to root the loop footpoints.

\subsection{\lya\ Light Curve and Comparison with Other Emissions}

\begin{figure}
\centering
\includegraphics[width=14cm]{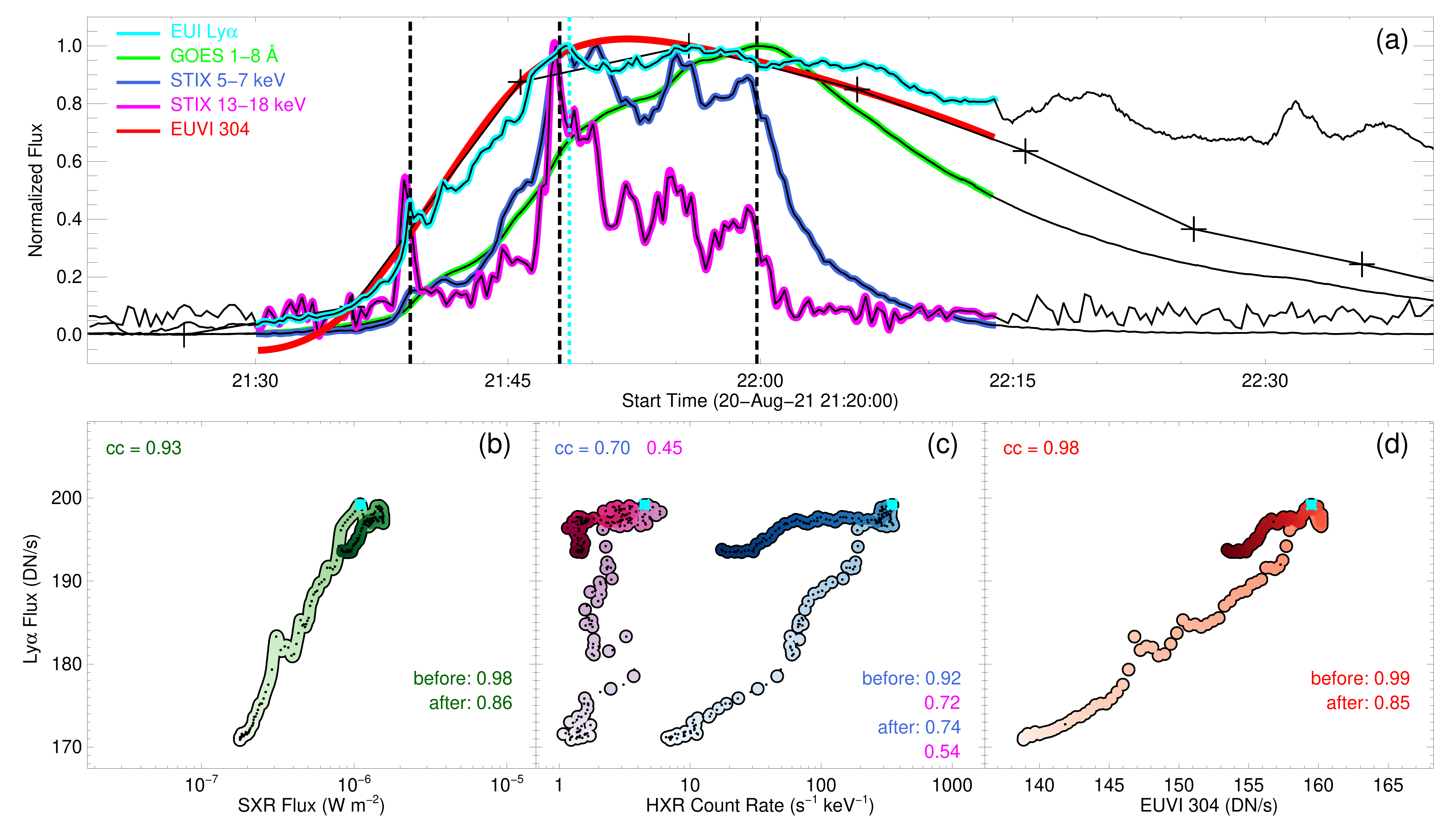}
\includegraphics[width=14cm]{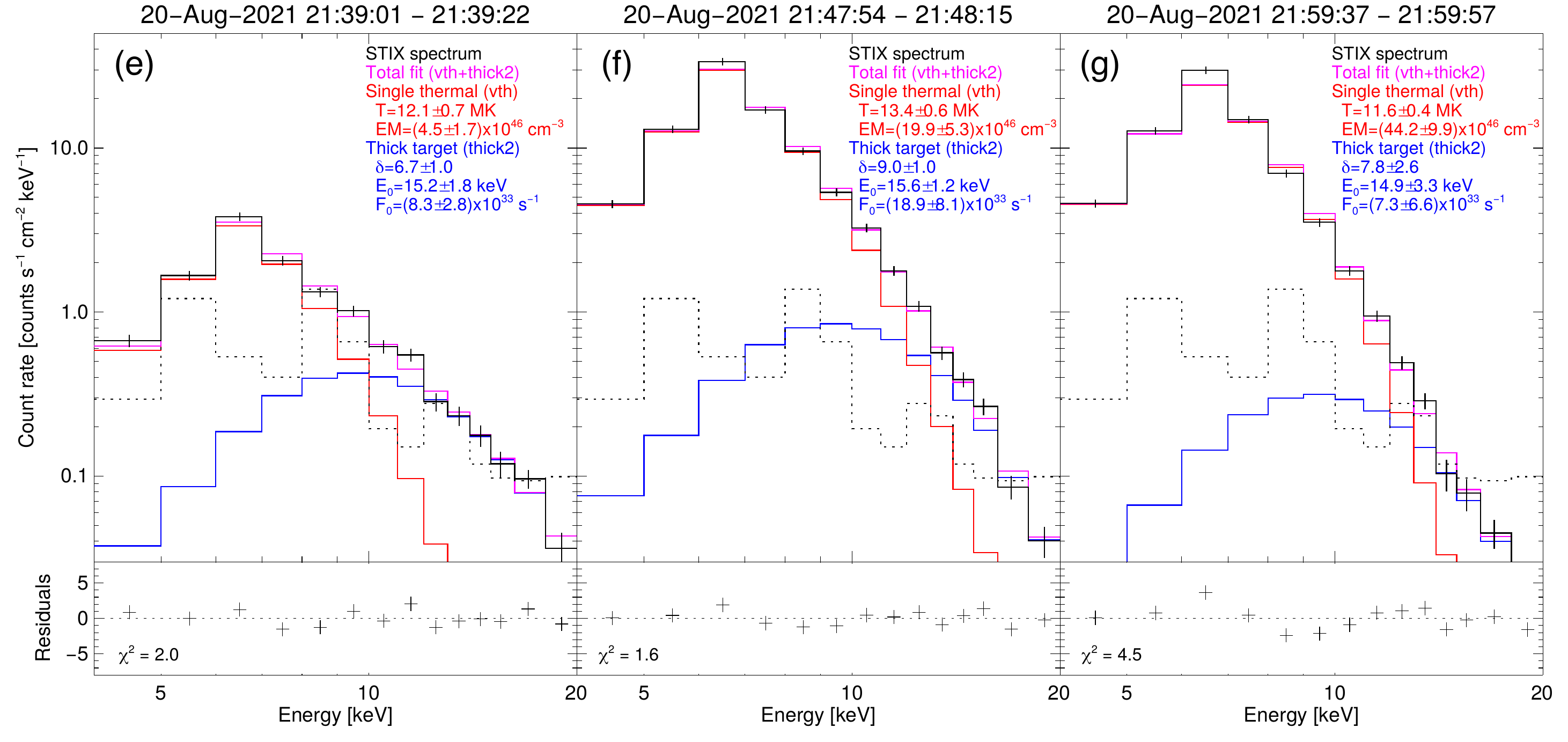}
\includegraphics[width=14cm]{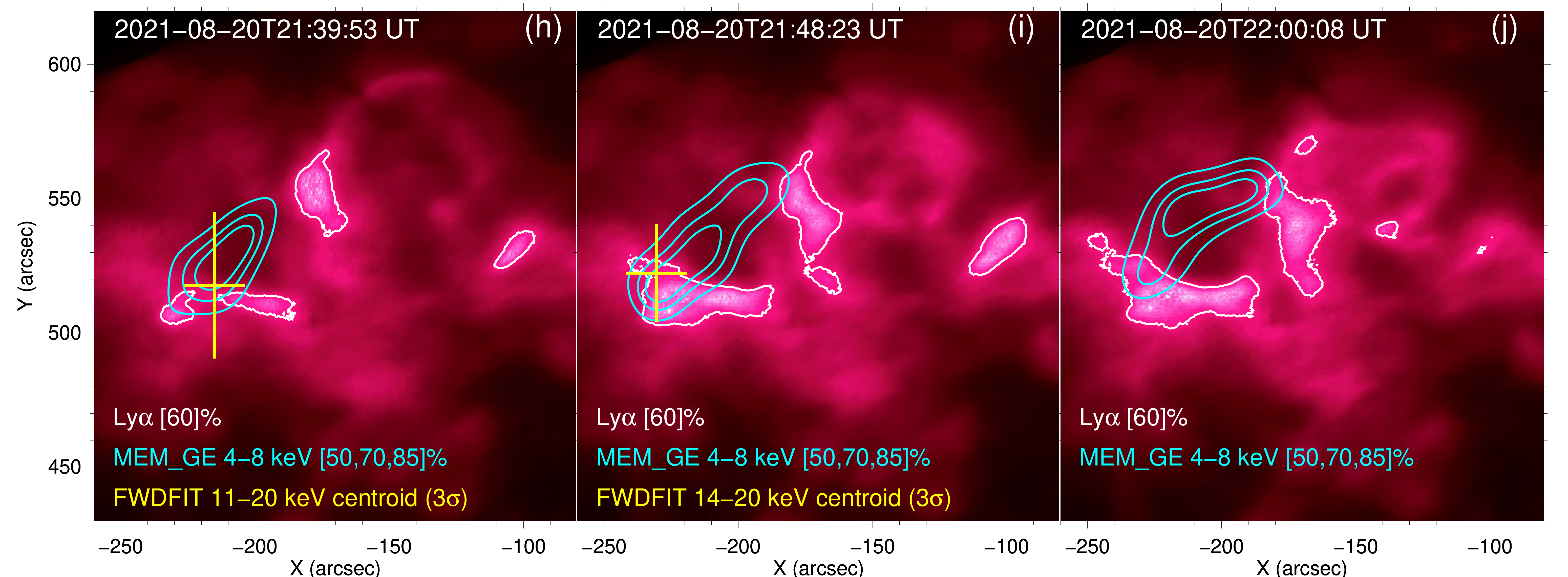}
\caption{(a) Cubic spline interpolations (color lines) to the background subtracted and normalized light curves from flare begin to end. (b)--(d) Scatter plots between the \lya\ emission and the other waveband emissions. The colors from light to dark represent the times from flare begin to end. The cyan star symbol denotes the \lya\ peak time (also indicated by the cyan vertical line in panel (a)). The  linear Pearson correlation coefficients are given, plus the ones before and after the \lya\ peak time. (e)--(g) HXR spectral fitting with a thermal plus nonthermal model for the three times marked by the black vertical lines in panel (a). The fitting parameters are shown in each panel. (h)--(j) HXR spectral imaging for the three selected times, with the thermal source contours (cyan color) overlaid on the \lya\ images. The yellow plus symbol shows the centroid of nonthermal sources.}
\label{fig:cc}
\end{figure}

From the \lya\ emission curve integrated over the whole flaring region as shown in Figure \ref{fig:lc}(b), we can see that it rises slowly, reaches its maximum at $\sim$21:48 UT (before the SXR peak time), and decays quite gradually. There are some small bumps or fluctuations either before the peak or after, which mainly originate from different flaring locations or structures (see Section \ref{sec:sub-spatio}) and do not show a good periodic property. When comparing with the SXR and HXR fluxes (Figures \ref{fig:lc}(a) and (b)), it is seen that the \lya\ emission has a similar peak time with the time derivative of SXR flux as well as with the HXR 5--7 keV and 13--18 keV fluxes (also see Figure \ref{fig:cc}(a)). In particular, the small bumps of \lya\ appearing before the flare peak time correspond to the peaks of the two HXR curves well. By comparison with the EUVI 304 \AA\ flux, \lya\ has a rise with similar slope but a flatter decay, at least for the whole flaring region. Owing to a low cadence, the peak time of EUVI 304 \AA\ is not quite clear but it should be similar with the one of \lya. Because of an occultation of flare ribbons plus a limb absorption by prominences or spicules, the AIA 304 \AA\ light curve (in Figure \ref{fig:lc}(c)) has a distinct shape with EUVI 304 \AA\ and thus with \lya. However, a careful inspection shows that the sub-peaks of AIA 304 \AA\ generally correspond to the small bumps of \lya\ during the entire flare period.

In order to investigate the relationship of the flaring \lya\ emission with the above waveband emissions quantitatively, we make scatter plots and calculate the linear Pearson correlation coefficients (cc) between them. Firstly, we subtract the minimum flux between 21:20--22:40 UT for each of the emission curves, which is assumed to be the background level. Note that most of the minimum fluxes appear at $\sim$21:25 UT, i.e., right before the flare onset. Secondly, we do cubic spline interpolations to the background subtracted curves for the time of 21:30--22:14 UT (i.e., from flare begin to end; see the color lines in Figure \ref{fig:cc}(a)) due to different sampling times. From the scatter plots in Figures \ref{fig:cc}(b)--(d), we can see that the \lya\ emission has good temporal relationships with the fluxes of SXR 1--8 \AA\ (cc$=$0.93), HXR 5--7 keV (cc$=$0.70), and 304 \AA\ (cc$=$0.98). It is interesting to find that there exist two distinct slopes before and after the \lya\ peak in the scattering plots (see the lighter and darker colors separated by the cyan star symbol). This is supposed to be caused by the distinct slopes of \lya\ rise and decay, the latter of which is much flatter. We also calculate the coefficients before and after the \lya\ peak separately and find even better relationships for the rise of \lya\ with the other emissions. 

We also explore the nonthermal/thermal nature of the flare and thus of \lya\ through HXR spectra. We first use a single thermal model plus a thick target model to do spectral fittings \citep{batt21} for three selected times that correspond to some HXR peaks, as plotted in Figures \ref{fig:cc}(e)--(g). It is seen that in the rise phase ($\sim$21:39 UT and $\sim$21:48 UT), some relatively prominent nonthermal signals show up, while the thermal component becomes more significant around the SXR peak time ($\sim$21:59 UT). The spectral fittings also indicate that the HXR 5--7 keV and 13--18 keV emissions as mentioned above represent the thermal and nonthermal components of the flare, respectively. We further reconstruct the HXR images for the three times and overlay HXR contours on \lya\ images in Figures \ref{fig:cc}(h)--(j). The MEM\_GE algorithm \citep{mass20} is used for the thermal sources in 4--8 keV, whereas for the nonthermal ones (above 11 keV) we only deduce the centroid location by means of the VIS\_FWDFIT algorithm \citep{hurf02} due to the low counting statistics. Preliminary STIX aspect solutions for the absolute location of the HXR sources have an uncertainty of $\sim$10\arcsec. We can see that the thermal sources shaping hot flare loops are connected by two \lya\ ribbons (i.e., conjugate footpoints), and the nonthermal centroids obtained for the rise phase are located at the eastern \lya\ ribbon. Note that not all the observed flare loop structures can be reconstructed in the HXR images.

\subsection{Spatio-temporal Variation of \lya\ along with Other Wavebands}
\label{sec:sub-spatio}

\begin{figure}
\centering
\includegraphics[width=18cm]{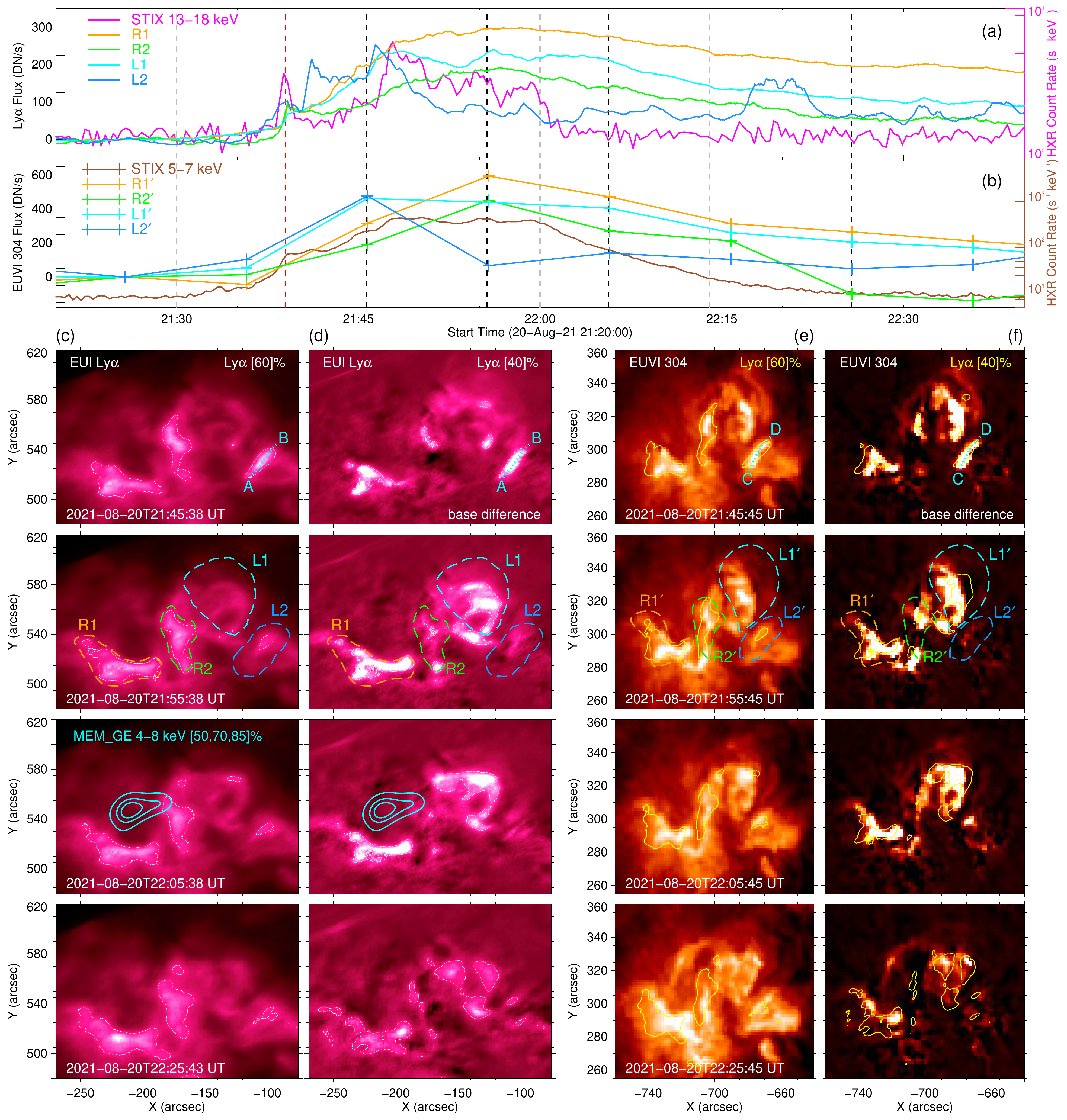}
\caption{(a) and (b) Base-image (at $\sim$21:25 UT) subtracted light curves of EUI \lya\ and EUVI 304 \AA\ from the four brightening kernels marked in panels (c)--(f), together with the STIX HXR fluxes. The three gray vertical lines indicate the flare onset, peak, and end times and the four black lines denote the four times of the \lya\ images in panels (c) and (d). The red vertical line marks the time of $\sim$21:39 UT when the \lya\ and HXR fluxes show a bump. (c) and (d) EUI \lya\ images and their base difference ones (with fixed flux scales) in different flare phases. The white contours in each frame represent the 60\% or 40\% level of \lya. The four enclosed dashed curves mark four brightening kernels, labelled as R1, R2, L1, and L2, whose light curves are plotted in panel (a). The cyan contours are the HXR thermal sources at the corresponding time. The slice AB is used to make a time-space diagram in Figure \ref{fig:slice}. (e) and (f) Similar to panels (c) and (d) but for EUVI 304 \AA. Note that the same \lya\ contours are overlaid on EUVI 304 \AA\ images with an alignment.}
\label{fig:img304}
\end{figure}

Thanks to the high-resolution imaging observations of EUI, we can investigate the spatio-temporal variation of the flaring \lya\ emission. Figure \ref{fig:img304}(c) displays the observed EUI \lya\ images in different flare phases. In order to remove the background or non-flaring source contributions, here we subtract the base image at $\sim$21:25 UT (i.e., right before the flare onset) from the observed images and show the base difference ones in Figure \ref{fig:img304}(d). It is seen that the \lya\ brightenings can come from both flare ribbons and loops. We select four brightening kernels (marked by four enclosed dashed curves in Figures \ref{fig:img304}(c) and (d)), two from ribbons (labelled as R1 and R2; connecting the thermal HXR 4--8 keV sources) and two from loops (labelled as L1 and L2), to make a further study. Their base-image subtracted light curves together with the HXR 13--18 keV flux are plotted in Figure \ref{fig:img304}(a). One can see that the \lya\ curves from the conjugate ribbons of R1 and R2 show a similar trend with a slow rise followed by a gradual decay. In particular, both curves display a small peak at $\sim$21:39 UT (indicated by the red vertical line), which matches the first peak of the nonthermal 13--18 keV emission. By contrast, the \lya\ curves from L1 and L2 exhibit multiple peaks at different times in either the rise or decay phase of the flare, most of which are also reflected in the \lya\ curve from the whole flaring region as shown in Figure \ref{fig:lc}(b). Note that L1 and L2 contain two loop systems with different magnetic connections and different length scales, which will be focused on in the following. Also note that the two \lya\ light curves from L1 and L2 show a peak or bump at $\sim$21:39 UT as well.

We also plot the corresponding EUVI 304 \AA\ images and their base difference ones in Figures \ref{fig:img304}(e) and (f) to make a comparison with \lya. Similarly, we select four brightening kernels, labelled as R1$^{\prime}$, R2$^{\prime}$, L1$^{\prime}$, and L2$^{\prime}$, corresponding to the ones in the \lya\ images. Note that the same \lya\ contours are overlaid on EUVI 304 \AA\ images with a co-alignment. One can see that the brightenings at 304 \AA\ match the ones of \lya\ very well except for some high-lying loops (say, from L1$^{\prime}$). This should be partly caused by a projection effect. In another word, the line-of-sight path of integration seems to be longer for these high-lying loops from the EUVI perspective, which makes them look brighter in EUVI 304 \AA\ images. The base-image subtracted light curves of 304 \AA\ from R1$^{\prime}$, R2$^{\prime}$, L1$^{\prime}$, and L2$^{\prime}$ are plotted in Figure \ref{fig:img304}(b). We can see that these 304 \AA\ curves show a similar trend with the corresponding \lya\ curves in general, although they have a very low cadence. 

For the four brightening kernels of R1, R2, L1, and L2, we also calculate the correlation coefficients of the flaring \lya\ emission with the other waveband emissions, as listed in Table \ref{tab:cc}. Note that a base image at $\sim$21:25 UT is subtracted for \lya\ and 304 \AA, and the SXR 1--8 \AA, HXR 5--7 keV, and 13--18 keV fluxes are the same with the ones in the flaring region case described above. From Table \ref{tab:cc}, we can see that the \lya\ emissions from R1, R2, and also L1 have good correlations with the fluxes of SXR 1--8 \AA, HXR 5--7 keV, as well as EUVI 304 \AA\ from R1$^{\prime}$, R2$^{\prime}$, and L1$^{\prime}$ (all cc$\ge$0.68). This result is similar to the flaring region case (with cc also listed in the table). However, L2 shows a different behaviour. Its \lya\ emission correlates a little better with the HXR 13--18 keV flux (cc$=$0.50) than the SXR 1--8 \AA\ and HXR 5--7 keV fluxes (cc$<$0.40), which will be discussed later.

\begin{table}[htb]
\caption{Correlation coefficients between the light curves of \lya\ and other wavebands}
\label{tab:cc}
\begin{tabular}{lcccc}
\hline
\hline
Location & \lya\ versus & \lya\ versus & \lya\ versus & \lya\ versus \\
         & ~~~SXR 1--8 \AA~~~~ & ~~~HXR 5--7 keV~~~ & ~~~HXR 13--18 keV~~~ & ~~~EUVI 304 \AA~~~~ \\
\tableline
Flaring Region & 0.93 & 0.70 & 0.45 & 0.98 \\
\hline
R1 & 0.96 & 0.68 & 0.39 & 0.99 \\
\hline
R2 & 0.96 & 0.78 & 0.49 & 0.97 \\
\hline
L1 & 0.94 & 0.74 & 0.48 & 0.95 \\
\hline
L2 & 0.16 & 0.39 & 0.50 & 0.88 \\
\hline
\hline
\end{tabular}
\end{table}

\begin{figure}
\centering
\includegraphics[width=15.5cm]{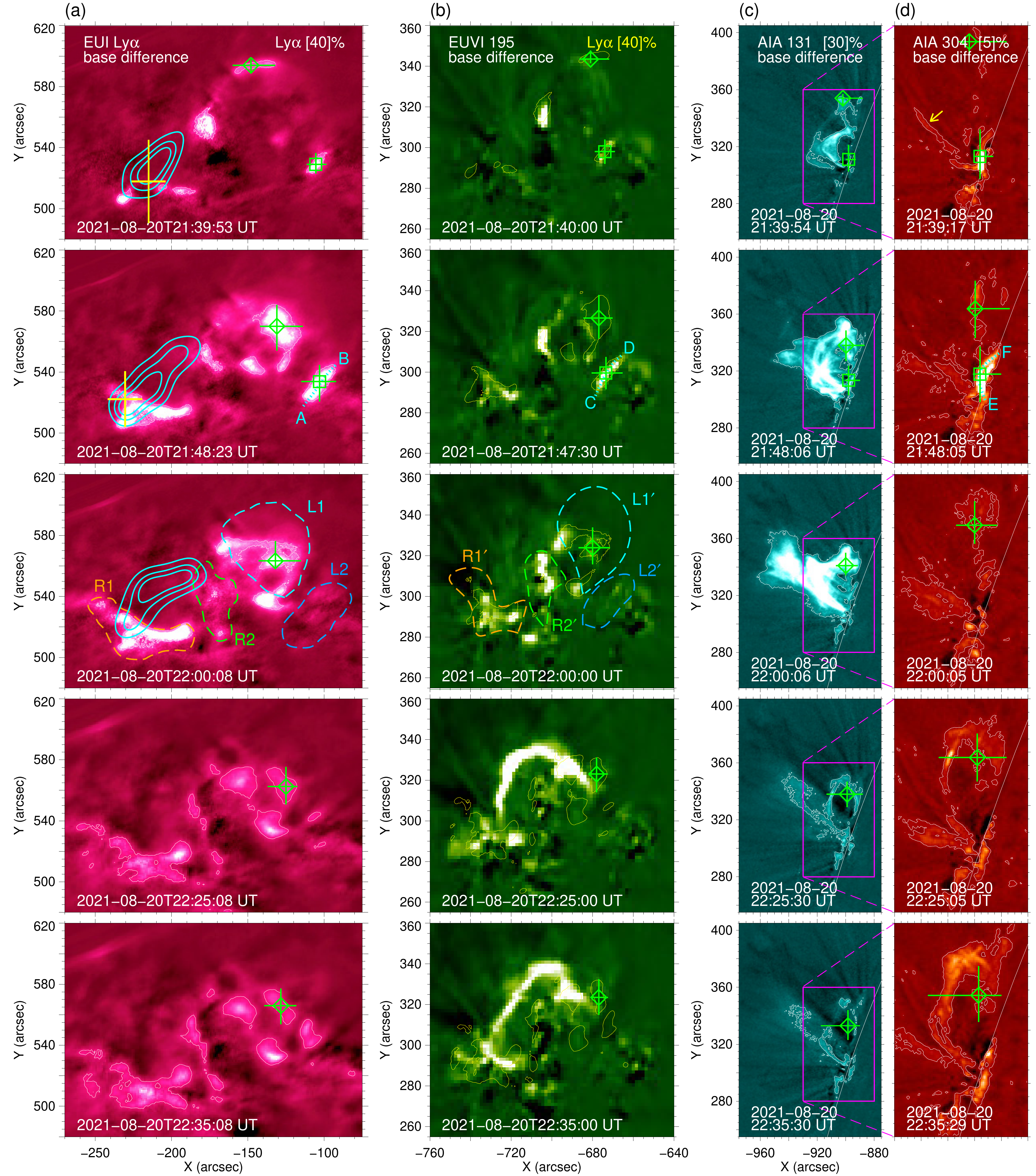}
\caption{Base difference images of EUI \lya\ (a), EUVI 195 \AA\ (b), AIA 131 \AA\ (c), and AIA 304 \AA\ (d) in different flare phases. The field of view of AIA 304 \AA\ is marked by the magenta box at AIA 131 \AA. The contours in each frame represent a certain flux level. Note that the same \lya\ contours with a 40\% level are overlaid on EUVI 195 \AA\ images. The green diamond and square symbols with an error denote the corresponding features as seen from three different perspectives. The cyan contours and yellow plus symbols in panel (a) are the same as the ones in Figures \ref{fig:cc}(h)--(j). The corresponding slices AB, CD, and EF are used to make time-space diagrams as shown in Figure \ref{fig:slice}. The brightening kernels of R1/R1$^{\prime}$, R2/R2$^{\prime}$, L1/L1$^{\prime}$, and L2/L2$^{\prime}$ are also indicated in panels (a) and (b) (see the four pairs of enclosed dashed curves). The yellow arrow in the first AIA 304 \AA\ image denotes coronal rain that can be clearly seen from sequential 304 \AA\ images. The images are available as an animation (top row for the observed images and bottom for the base-difference ones) which presents a temporal evolution of the C1.4 flare in EUI \lya\ (with a cadence of 15 or 5 s), EUVI 195 \AA\ (with a cadence of 2.5 min), AIA 131 \AA\ (with a cadence of 12 s) and AIA 304 \AA\ (with a cadence of 12 s) from 21:09 UT to 22:49 UT on 2021 August 20. The real-time duration of the animation is about 22 s.}
\label{fig:img195}
\end{figure}

We show more base difference images of \lya\ along with the ones of EUVI 195 \AA, AIA 131 \AA, and AIA 304 \AA\ in Figure \ref{fig:img195} (also see the accompanying animation) to focus on the \lya\ emissions from the flare loops at L1 and L2. Firstly, the bright loop structures from L1 and L2 are confirmed by a 3D reconstruction via $scc\_measure.pro$ which can provide corresponding pixels from the images with different perspectives \citep{inhe06}. The corresponding bright features between \lya\ and 304 \AA\ from L1/L1$^{\prime}$ and L2/L2$^{\prime}$ are identified and indicated by the green diamond and square symbols with an error in the images. Resorting to the side view of AIA, we ensure that these notable brightenings mainly come from flare loops, particularly from low-lying loops marked by the square symbols. In addition, from the AIA 131 \AA\ images (Figure \ref{fig:img195}(c)) we can clearly see various loop systems including the ones from L1/L1$^{\prime}$ and L2/L2$^{\prime}$ and also the main flare loops connected by R1/R1$^{\prime}$ and R2/R2$^{\prime}$, which show up at different times. The large-scale main flare loops should correspond to the thermal sources at HXR 4--8 keV (see the HXR contours in Figure \ref{fig:img195}(a)) and have a high temperature (say, log($T$)$\ge$6.8). Thus they cannot be seen in \lya\ that is sensitive at low temperatures. However, the loops without HXR counterparts, i.e., from L1/L1$^{\prime}$ and L2/L2$^{\prime}$, are probably filled with low temperature plasmas and can be seen in \lya, 304 \AA, and also 131 \AA\ (having a low temperature component). Note that the hot main flare loops cool down gradually and can be well seen in EUVI 195 \AA\ images (Figure \ref{fig:img195}(b)) at later times (say, 22:25 UT and 22:35 UT).

\begin{figure}
\centering
\includegraphics[width=18cm]{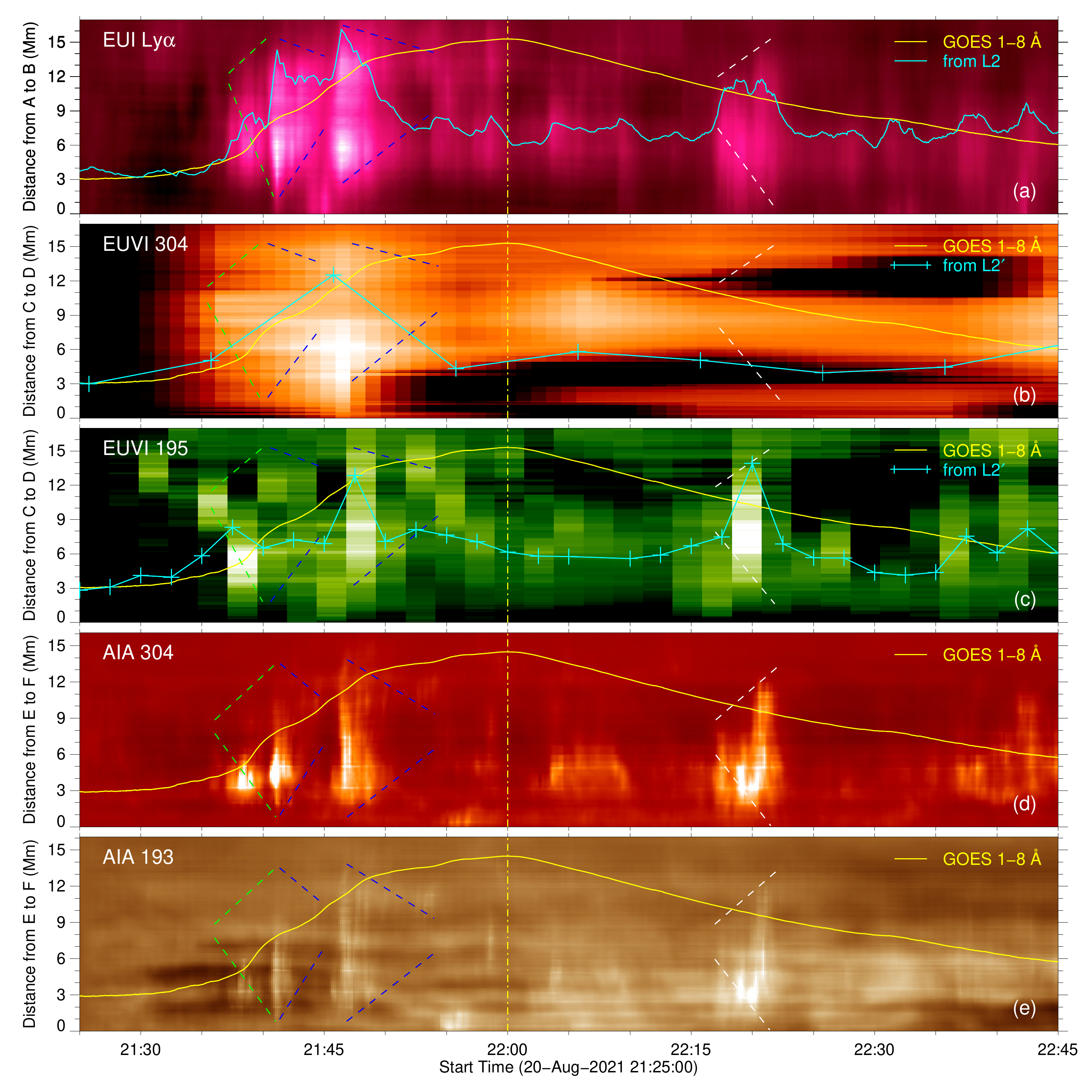}
\caption{Time-space diagrams of slices AB, CD, and EF (marked in Figures \ref{fig:img304} and \ref{fig:img195}) obtained from base difference images. The light curves of SXR 1--8 \AA\ and EUI \lya, EUVI 304 \AA, and 195 \AA\ from L2/L2$^{\prime}$ are overlaid on the panels. The vertical dash-dotted line denotes the flare peak time. Some pairs of dashed lines indicate possible plasma motions along the flare loops.} 
\label{fig:slice}
\end{figure}

Furthermore, we cut three corresponding slices (denoted by AB, CD, and EF in Figures \ref{fig:img304} and \ref{fig:img195}) along the low-lying loops from L2/L2$^{\prime}$ and plot their time-space diagrams (obtained from base difference images) in Figure \ref{fig:slice}. One can see that the \lya\ brightenings appear not only in the rise phase but also in the decay phase, which correspond to the brightenings at EUVI 304 \AA, 195 \AA, AIA 304 \AA, and 193 \AA\ generally. Note that the emissions of AIA 304 \AA\ and especially 193 \AA\ from such low-lying loops suffer a limb absorption by relatively cool spicules and thus several dark stripes show up in the diagram (say, at 2 Mm and 5 Mm of the slice between 21:30 UT and 21:45 UT). One could also see some promising plasma motions along the loops (marked by a few pairs of dashed lines) with speeds of several tens of km s$^{-1}$. Some plasmas are moving from the loop-top downward to legs or footpoints (denoted by the green and white pairs), while some have an opposite direction (see the blue pairs).

\section{Summary and Discussions}
\label{sec:summary}

In this study, we present novel \lya\ images plus HXR spectral observations from SO/EUI and SO/STIX, combining with multi-waveband and multi-perspective observations from STEREO-A/EUVI and SDO/AIA, to study the physical origin and property of the \lya\ emission in a C1.4 flare. To our knowledge, this is the first reported flare study on EUI \lya. The observational results are summarized below.

\begin{enumerate}
\item The flaring \lya\ emission has good  temporal correlations with the thermal emissions of SXR 1--8 \AA\ and HXR 5--7 keV.
\item The centroids of nonthermal HXR sources ($>$11 keV) obtained in the rise phase of the flare are co-spatial with the \lya\ brightening ribbons. 
\item The brightened \lya\ emissions can originate from flare ribbons as well as flare loops (especially cool low-lying loops), whose spatio-temporal variations are fairly similar with the ones at EUVI 304 \AA.
\item The \lya\ brightenings from the flare loops can appear not only in the rise phase but also in the decay phase of the flare, which correspond to the ones at 304 \AA\ and 195 \AA\ generally. 
\end{enumerate}

\subsection{Physical Origin of the Flaring \lya\ Emission}

Firstly, the good temporal relationships of the flaring \lya\ emission (either from ribbons or at loops) with the thermal emissions of HXR 5--7 keV as well as SXR 1--8 \AA\ in this C1.4 flare suggest that the flaring \lya\ emission is mainly produced by a thermal process, say thermal conduction heating and cooling. This could also explain the slow rise and flat decay of \lya\ more easily. Meanwhile, some spiked bumps of \lya\ that correspond to the HXR 13--18 keV peaks during the rise phase of the flare can be produced by nonthermal electrons. These \lya\ emissions mainly originate from flare ribbons. The \lya\ brightenings from flare loops are more likely caused by plasma heating and also cooling occurring in different flare phases. These could be accompanied by some plasma motions such as chromospheric evaporation and coronal condensation, which have different directions along the loops. Compared with previous studies on flaring \lya\ \citep[e.g.,][]{nusi06,rubi09,jing20}, our study provides a direct imaging evidence for the \lya\ emission originating from flare ribbons as well as flare loops for the first time. In addition, the thermal origin of \lya\ is supposed to be more important than the nonthermal origin in this small flare, which may be different with the case in large or major flares.

A more detailed possible scenario for flaring \lya\ could be speculated from the comprehensive observations of this flare (see the accompanying animation of Figure \ref{fig:img195}). Before the flare onset, pre-existing coronal rains (also shown in AIA light curves in Figure \ref{fig:lc}(c)) from large-scale loops fall down and interact with the low-lying magnetic loops at L2/L2$^{\prime}$ (as clearly seen in the animation), which leads to a magnetic reconnection and triggers the C1.4 flare. This process proceeds into the rise phase of the flare (also see the first AIA 304 \AA\ image in Figure \ref{fig:img195}(d)) and leads to a further magnetic reconnection. The magnetic reconnection takes place in different loop systems that have different magnetic connections and different length scales (i.e., various loops as seen in AIA 131 \AA\ images), which results in plasma heating as well as electron acceleration during the rise phase. Thus we see the \lya\ brightenings at L2, L1, as well as R1 and R2 simultaneously (i.e., at $\sim$21:39 UT and $\sim$21:48 UT), accompanied by nonthermal HXR 13--18 keV emissions co-spatial with the flare ribbons (say, R1). This could also explain the temporal relationship between the \lya\ emission from L2 and the HXR 13--18 keV flux. Some (probably most) of the \lya\ emissions at the ribbons may also be generated by thermal conduction heating from the reconnected or flare loops, therefore, the \lya\ emissions from R1 and R2 have a good relationship with the SXR 1--8 \AA\ and HXR 5--7 keV emissions coming from hot flare loops. Note that the loops at L1 are connected by the ribbon of R2 (at least for one end), so its \lya\ emission also has a good temporal correlation with the SXR 1--8 \AA\ and HXR 5--7 keV emissions, just like the emission at R2. As for the \lya\ brightenings appearing in the decay phase of the flare, coronal rain falling (due to plasma cooling) and also impacting on the solar surface (at the flare ribbons) are supposed to play some roles. This could probably explain the very gradual decay of \lya\ to some extent, as well as the brightening patches of \lya\ at the flare ribbons of R1 and R2 at late times.

It should be mentioned that the broadband nature of EUI HRI images has some influences on the observed \lya\ emission of this waveband. Two interference filters were used to achieve the spectral purity of \lya, one with 200 \AA\ FWHM bandwidth and the other with 100 \AA\ bandwidth \citep{roch20}. Note that the narrow band filter has a peak response at 1190 \AA. Therefore, some other lines such as Si {\sc iii} at 1207 \AA\ and S {\sc x} at 1196 and 1213 \AA\ \citep{curd01,curd04} and the continuum nearby could contribute to this waveband emission, particularly during the flare. They may smooth out the \lya\ emission curve \citep{mill16} and thus lead to a relatively slow rise and also a gradual decay of \lya. In addition, the ``remnant image effect" of the HRI \lya\ telescope causing subsequent images to show a fraction of the previous images proportional to the actual incoming intensity is another potential factor that could contribute to a gradual decay of \lya. Considering that some bright regions (say, kernel L2) can decay very quickly (in a few minutes) and that the \lya\ brightenings are quite consistent with the ones of 304 \AA\ in spatio-temporal variation, the remnant image effect in \lya\ may be trivial for the flare under study. More validation works for the remnant image effect of EUI HRI can be found in Appendix \ref{app:lya} via a comparison between the \lya\ emissions observed by EUI and the EUV Sensor (EUVS; \citealt{vier07}) aboard GOES for two flares (see Figure \ref{fig:app-lya}). The comparison between the EUVS \lya\ and AIA 304 \AA\ emissions from five flares (Figure \ref{fig:app-304}) in Appendix \ref{app:304} also indicates that the relatively flatter decay of \lya\ mainly originates from the flaring plasma itself rather than the remnant image effect of EUI HRI.

\subsection{Relationship between \lya\ and He {\sc ii} 304 \AA\ during the Flare}

The flaring \lya\ emission, either at the ribbons or from the loops, shows similar features with the emission of He {\sc ii} 304 \AA\ (the \lya\ transition of He {\sc ii}) in the light curve as well as spatio-temporal variation. This study provides a direct imaging evidence for the similarities of these two waveband emissions during flares by comparison with prior studies that usually focus on the light curves and radiated energy. For instance, \cite{mill12} found that the \lya\ and He {\sc ii} 304 \AA\ lines exhibit a similar trend in the integrated intensity and also emit a comparable energy within an order of magnitude in an X-class flare. \cite{kret13} compared the intensity curves of the two lines and found an excellent match including the amount of radiated energy. On the other hand, some small differences between the two waveband emissions could also be seen in the studied flare, say, a flatter decay of the \lya\ light curve and brighter flare loops in the 304 \AA\ images. Note that the relatively flatter decay of \lya\ can also be found in some other flare events by comparing the full-Sun EUVS \lya\ and AIA 304 \AA\ emissions (see Figure \ref{fig:app-304} in Appendix \ref{app:304}). These differences may be explained by slightly different formation layers (including the continuum within the two wavebands) or opacity effects of the two lines. A further study is worthwhile particularly combining with radiative hydrodynamic simulations in the future.

\subsection{Insights to the Future \lya\ Work}

To investigate the physical origin of flaring \lya, we combine EUI \lya\ images with STIX HXR spectra, the latter of which can provide the information on hot thermal plasmas as well as accelerated nonthermal electrons. More flare events, especially large ones, need to be studied in the future. In addition, multi-perspective observations are quite necessary to fully understand the flaring \lya\ emission. The Chinese Advanced Space-based Solar Observatory (ASO-S; \citealt{ganw19}) carries a \lya\ Solar Telescope \citep{lihu19} that can provide the full-Sun \lya\ images and also a HXR Imager \citep{zhan19} observing the Sun in the energy range of $\sim$30--200 keV. ASO-S will be launched into a Sun-synchronous orbit in late 2022. Potential joint observations of ASO-S and SO will benefit us to study the \lya\ emission in a more extensive way \citep{vour19,kruc19}.

\appendix

\section{Comparison between EUI and EUVS \lya\ Emissions from Two Flare Events}
\label{app:lya}

\begin{figure}
\centering
\includegraphics[width=14cm]{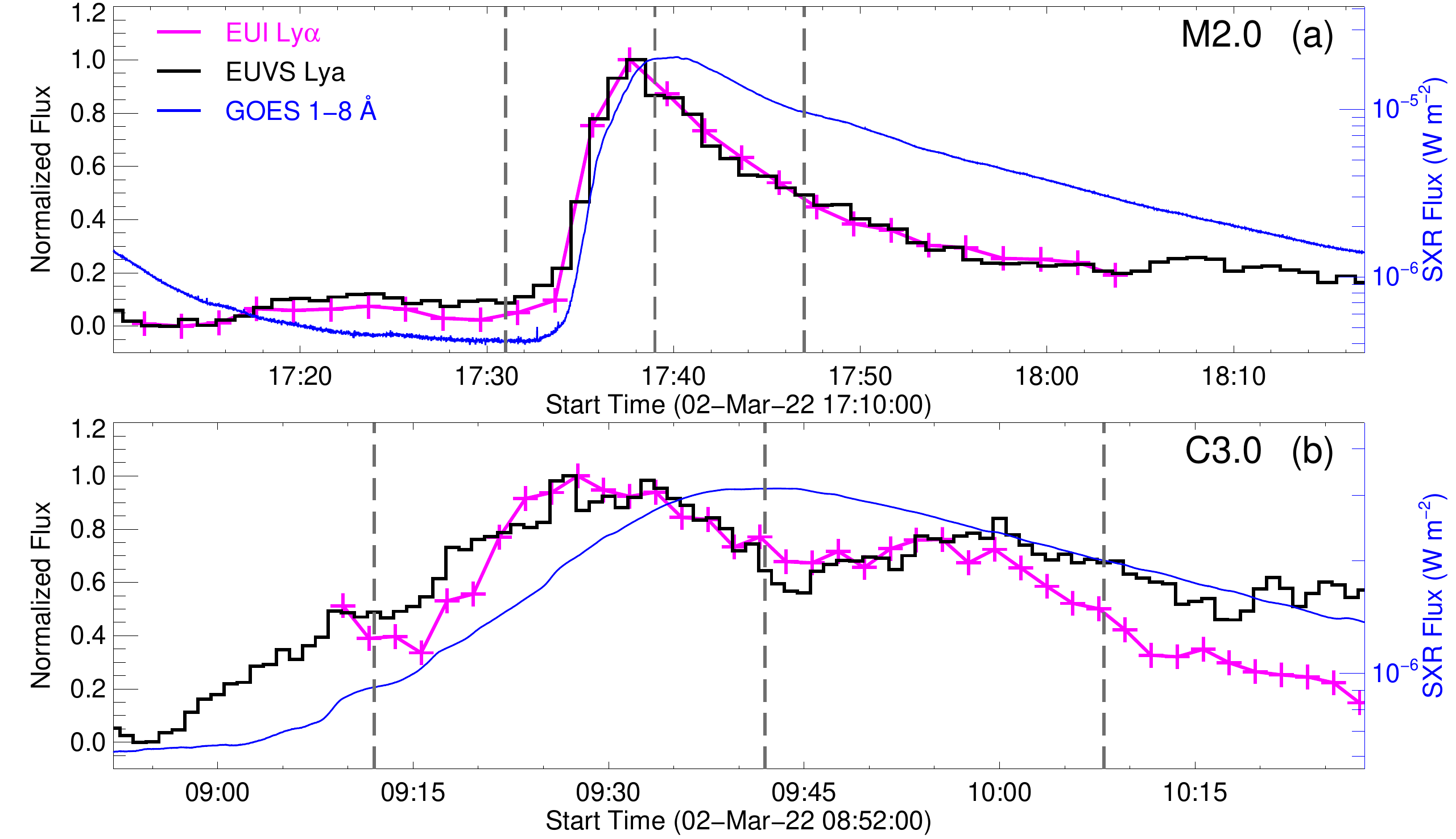}
\caption{Comparison between the normalized EUI and EUVS \lya\ emissions from two flare events (an M2.0 (a) and a C3.0 (b)) on 2022 March 2. Note that the former one is obtained by integrating over the whole field of view of EUI and the latter one is from the full Sun. The three vertical dashed lines indicate the onset, peak, and end times of the flare.}
\label{fig:app-lya}
\end{figure}

We compare the \lya\ emission curves observed by EUI and EUVS for two flares, an M2.0 and a C3.0 on 2022 March 2, as shown in Figure \ref{fig:app-lya}. On that day the two instruments observed the Sun nearly in the same perspective, i.e., along the Earth-Sun connection line. Note that the EUVS \lya\ emission is from the full Sun and the EUI \lya\ curve is obtained by integrating over the whole field of view (about one sixth of the full disk). For the large M2.0 flare (Figure \ref{fig:app-lya}(a)), we can see that the normalized EUI \lya\ curve matches the one of EUVS very well. Regarding the small C3.0 flare (Figure \ref{fig:app-lya}(b)), the EUI \lya\ curve shows a similar trend as the one of EUVS in the rise phase of the flare, while it decays more quickly during the decline phase. This difference is very likely caused by different integration regions of the \lya\ emissions from EUI and EUVS, which can be significant for small flares. The overall match of EUI and EUVS  \lya\ emissions  may imply that the remnant image effect of EUI HRI plays a minor role in the decay of \lya\ during flares.

\section{Comparison between EUVS \lya\ and AIA 304 \AA\ Emissions from Some Flare Events}
\label{app:304}

\begin{figure}
\centering
\includegraphics[width=14cm]{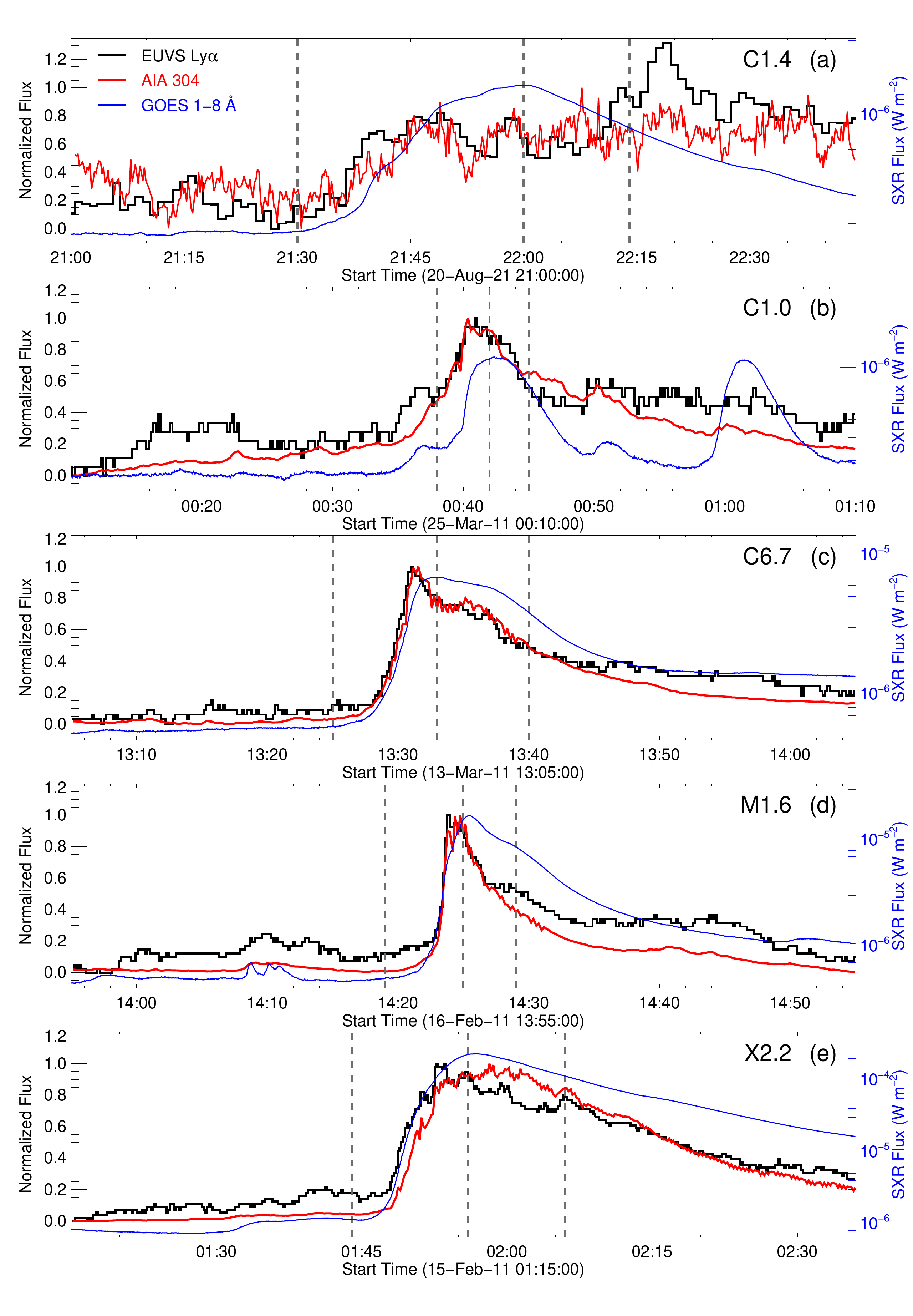}
\caption{Comparison between the normalized full-Sun EUVS \lya\ and AIA 304 \AA\ emissions for five flares (from C- to X-class ones including the C1.4 flare under study). The three vertical dashed lines indicate the onset, peak, and end times of the flare.}
\label{fig:app-304}
\end{figure}

We plot the full-Sun EUVS \lya\ and AIA 304 \AA\ emission curves for five flares (from small to large ones including the C1.4 flare under study) to compare the decay of \lya\ and 304 \AA\ in Figure \ref{fig:app-304}. Note that all these curves are observed from the Earth perspective. It is seen that the \lya\ emission shows a relatively flatter decay than 304 \AA\ in these flares. In particular, for the C1.4 flare studied here (Figure \ref{fig:app-304}(a)), the \lya\ emission is notably enhanced in the late decay phase (during $\sim$21:15--21:25 UT, corresponding well to the brightening from L2) when the flaring region was rotating onto the visible solar disk), while the 304 \AA\ emission just increases slightly then. All these indicate that the \lya\ emission during solar flares usually has a relatively flatter decay than the 304 \AA\ emission.

\acknowledgments
We thank the anonymous referee very much for telling us the ``remnant image effect" of the EUI HRI \lya\ telescope as well as providing detailed suggestions/comments that help us to improve the manuscript. The data are courtesy of the teams of SO, STEREO, SDO, and GOES. Solar Orbiter is a space mission of international collaboration between ESA and NASA, operated by ESA. The EUI instrument was built by CSL, IAS, MPS, MSSL/UCL, PMOD/WRC, ROB, LCF/IO with funding from the Belgian Federal Science Policy Office (BELSPO/PRODEX PEA C4000134088); the Centre National d'Etudes Spatiales (CNES); the UK Space Agency (UKSA); the Bundesministerium f{\"u}r Wirtschaft und Energie (BMWi) through the Deutsches Zentrum f{\"u}r Luft- und Raumfahrt (DLR); and the Swiss Space Office (SSO). STIX is supported by SSO (the lead funding agency for STIX), PNSC, CNES, CEA, CME (via the PRODEX program), DLR, ASP, ESA PRODEX, ASI and INAF. SDO is a mission of NASA’s Living With a Star Program. We thank Dr. Beili Ying, Jianchao Xue, and Xiaoyu Xiafan for the help of data processing. The authors are supported by NSFC under grants 11873095, 11733003, 11921003, and 11961131002, and by the CAS Strategic Pioneer Program on Space Science under grants XDA15052200, XDA15320103, and XDA15320301. Y.L. is also supported by the CAS Pioneer Talents Program for Young Scientists.

\bibliographystyle{apj}

\end{document}